\documentclass[runningheads]{llncs}

\AtBeginDocument{%
  }

\usepackage{tikz}
\usetikzlibrary{arrows.meta, positioning, calc}
\usepackage{graphicx}
\usepackage{algorithm}
\usepackage{algpseudocode}
\usepackage{pslatex}
\usepackage{amsmath}
\usepackage{booktabs}
\usepackage{makecell}

\usepackage{pgfplots}
\usepackage{subcaption}
\usepackage{booktabs}
\usepackage{xcolor}
\usepackage{hyperref}

\pgfplotsset{compat=1.18}

\usepackage{multirow}
\definecolor{nvidiagreen}{HTML}{76B900}
\definecolor{amdred}{HTML}{ED1C24}
\definecolor{intelblue}{HTML}{0071C5}

\definecolor{gatkred}{HTML}{ED1C24}
\definecolor{gatksparkorange}{HTML}{FF5F15}
\definecolor{syclgalblue}{HTML}{0071C5}
\definecolor{parabricksgreen}{HTML}{76B900}

\title{X-BQSR: Holistic Acceleration of Base Quality Recalibration for Scalable Genomic Analysis}
\titlerunning{X-BQSR}

\author{Eugenio Marinelli\thanks{These authors contributed equally to this work.} \and
Ivan Donchev Kabadzhov\textsuperscript{$\star$} \and
Raja Appuswamy}
\authorrunning{Eugenio Marinelli, Ivan Donchev Kabadzhov and Raja Appuswamy}
\institute{EURECOM, Biot, France}

\begin{document}

\maketitle

\begin{abstract}
  With the continued decline in sequencing costs, genomics-based precision medicine is gaining increasing importance. The Genome Analysis Toolkit (GATK) is the standard framework for converting raw genomic data into actionable variants, but one of its most expensive steps is base quality score recalibration (BQSR), which corrects systematic sequencing errors. Despite significant work on accelerating other parts of the variant calling pipeline, BQSR has not been systematically explored on GPUs.

In this work, we analyze BQSR to identify computational and I/O bottlenecks and show that simply offloading key stages to an accelerator is insufficient to achieve overall gains. We then present X-BQSR, a holistic redesign of the entire BQSR pipeline, including I/O, compression, and kernel execution, to better leverage GPU parallelism. We implement X-BQSR using portable SYCL kernels for NVIDIA, AMD, and Intel GPUs, along with native CUDA and HIP backends for comparison, and we evaluate its performance against previous FPGA work. Our evaluation shows up to 141.7× acceleration compared to standard GATK and 8.7× acceleration compared to a proprietary GPU-accelerated solution.
\end{abstract}

\section{Introduction}

\begin{figure}[t]
    \centering
    \includegraphics[width=0.80\linewidth]{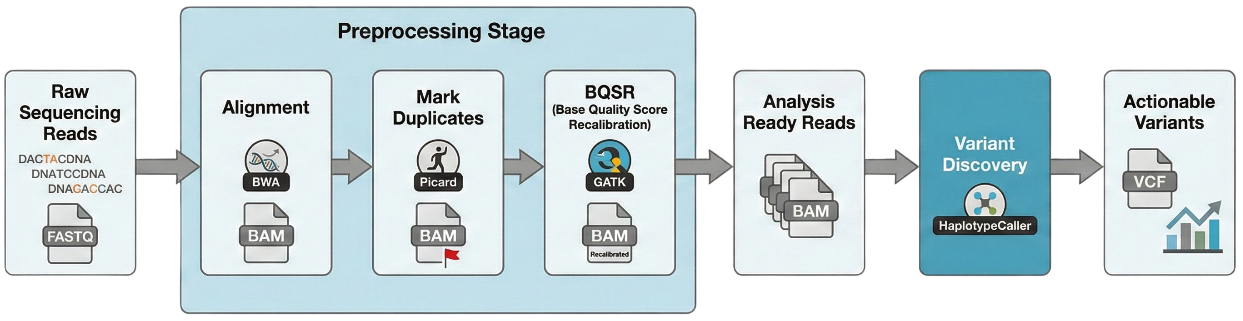}
    \caption{GATK germline variant calling pipeline.}
    \label{fig:gatk-germline-pipeline}
    \vspace{-5mm}
\end{figure}

Over the last two decades, the cost of DNA sequencing has dropped faster than Moore's law, from over \$1,000,000 in 2005 to well under \$1,000 today~\cite{SequencingCost}. As we move into the era of \$100 sequencing, genomics-driven precision medicine efforts have become widespread in both research and clinical practice. Germline genomic analysis, or the task of studying mutations that are inherited and present in every cell, is of vital importance as it has several applications, from clinical diagnosis of rare diseases in humans, to population-genomics studies that map genetic diversity in other organisms. The gold standard computational pipeline used for germline genomic analysis is the GATK~\cite{McKenna2010GATK,DePristo2011,GATK42019} germline variant calling pipeline shown in Figure~\ref{fig:gatk-germline-pipeline}.

It is well known that the GATK pipeline is computationally intensive. For instance, an evaluation by Intel~\cite{36-cores-Prabhakaran2017AccelerateGR} showed that the pipeline took 1.5 days to run the pipeline for a whole-genome sequencing dataset at 30$\times$ coverage on a 36-core CPU. This naturally makes the pipeline a key bottleneck in any genomic analysis that processes multiple such sequences, and hence, a key optimization target for hardware-accelerated solutions from both industry and academia. On the research front, several solutions have been proposed to accelerate the sequence alignment and variant discovery stages of the GATK pipeline shown in Figure~\ref{fig:gatk-germline-pipeline} using CPU with highly optimized parallel processing and analysis tools~\cite{elprep2015}, GPUs~\cite{oconnell2023parabricks,soap3,soap3dp,barracuda,cushaw}, FPGAs~\cite{fpga-bqsr,hybrid-sra}, and other custom accelerators~\cite{acceleratorGATK}.

However, to our knowledge, the base quality score recalibration (BQSR) stage has been the focus of only one academic study, which proposed an algorithm-hardware co-optimization strategy using FPGA by Lo et al~\cite{fpga-bqsr}. While that work demonstrated the potential of hardware acceleration, it only targeted the BaseRecalibrator phase of BQSR, neglected the ApplyBQSR phase, and did not perform end-to-end integration within the full GATK pipeline.

Furthermore, there is currently no prior work in the literature systematically investigating the challenges of adapting the BQSR algorithm to exploit the SIMT parallelism of GPUs. While the NVIDIA Parabricks library provides a high-performance commercial solution for BQSR acceleration~\cite{Parabricks2020}, it is a closed-source binary. This opacity prevents the research community from inspecting its underlying strategies, and its CUDA-based nature lacks portability to multi-vendor accelerators, such as Intel Flex or AMD Instinct GPUs.

In this work, we bridge this gap by making several contributions. First, we present an in-depth analysis of BQSR to identify various computational and I/O bottlenecks in the CPU-based implementation and to identify GPU-offload targets for an accelerated redesign. Second, we present \textbf{X-BQSR}, a redesign of the core BQSR algorithms to unlock the massive SIMT parallelism of GPUs. In doing so, we not only develop optimized GPU kernels for performing the core computation but also perform a holistic redesign of all other aspects of BQSR including I/O, compression, memory management, and data transfers, to ensure that the improvement unlocked by GPU acceleration is not overshadowed by other hidden overheads. Third, we implement three variants of \textbf{X-BQSR} using CUDA, HIP, and SYCL. Our  SYCL \textbf{X-BQSR} implementation is the first, portable, vendor-neutral, open-source implementation of BQSR that can run on multicore CPUs, integrated GPUs, and multi-vendor discrete GPUs using a single code base. Finally, we present a rigorous evaluation comparing \textbf{X-BQSR} with the original single-threaded GATK BQSR, Spark-enabled multi-threaded GATK BQSR, FPGA BQSR from Lo et al., and NVIDIA Parabricks BQSR. In doing so, we show that: (i) \textbf{X-BQSR} can outperform NVIDIA Parabricks by $3.6\times$ to $8.7\times$ across varying dataset scales; (ii) SYCL \textbf{X-BQSR} matches or exceeds the performance of its CUDA counterpart while remaining portable across NVIDIA, AMD, and Intel GPUs; and (iii) unlike FPGA implementations that require external data pre-formatting, our implementation serves as a drop-in replacement in the GATK germline pipeline, providing up to a $141.7\times$ speedup at the BQSR level and a $2.7\times$ speedup for the GATK preprocessing stage as a whole compared to optimized multithread CPU workflows. We plan to make our code fully open source together with the camera-ready version of our manuscript.
\section{Design}
\label{sec:design}

\begin{table}[t]
\small
\centering
\caption{Pipeline and BQSR breakdown for HG004 (NovaSeq)}
\label{tab:pipeline_breakdown}
\begin{minipage}[t]{0.52\textwidth}
\centering
\caption*{(a) Pipeline-level breakdown}
\begin{tabular}{lrr}
\toprule
Stage & \makecell[r]{Time\\(min)} & \makecell[r]{Share\\(\%)} \\
\midrule
Alignment (bwa-mem2) & 156.0 & 18.9 \\
GATK Sort / Markdup & 219.0 & 26.5 \\
\makecell[l]{BQSR model build\\(\texttt{BaseRecalibrator})} & 223.8 & 27.1 \\
\makecell[l]{BQSR apply\\(\texttt{ApplyBQSR})} & 226.8 & 27.5 \\
\bottomrule
\end{tabular}
\end{minipage}
\hfill
\begin{minipage}[t]{0.44\textwidth}
\centering
\caption*{(b) BQSR workflow breakdown}
\label{tab:bqsr_breakdown}
\begin{tabular}{lrr}
\toprule
Component & \makecell[r]{Time\\(min)} & \makecell[r]{Share\\(\%)} \\
\midrule
\textbf{Input Processing} & \textbf{68.46} & \textbf{32.3} \\
\quad Parsing & 62.71 & 29.6 \\
\quad Decompression & 5.56 & 2.6 \\
\addlinespace[0.3em]
\textbf{Computation} & \textbf{58.14} & \textbf{27.4} \\
\quad \texttt{BaseRecalibrator} & 38.63 & 18.2 \\
\quad \texttt{ApplyBQSR} & 19.51 & 9.2 \\
\addlinespace[0.3em]
\textbf{Output Processing} & \textbf{85.50} & \textbf{40.3} \\
\quad Formatting & 27.50 & 13.0 \\
\quad Compression & 56.98 & 26.8 \\
\bottomrule
\end{tabular}
\end{minipage}
\vspace{-5mm}
\end{table}

Modern high-throughput sequencing (HTS) instruments read DNA one base at a time and  
assign a quality score to each base call that is intended to approximate the probability that the predicted base is incorrect. While instrument calibration provides a reasonable initial estimate, systematic biases commonly remain that lead to errors in estimates. These errors may depend on sequencing context (e.g., local k-mer patterns), cycle (position within the read), and lane-specific effects. Since downstream variant callers weight evidence using base qualities, miscalibration can translate into misweighted observations at scale, affecting the accuracy of variant calls~\cite{Ewing1998Phred,DePristo2011}.
BQSR calibrates base-quality scores by estimating empirical error rates conditioned on covariates while excluding known-variant loci. The workflow has two stages: (1) model construction (\texttt{BaseRecalibrator}), which takes as input a BAM file containing aligned reads and builds recalibration statistics, and (2) score application (\texttt{ApplyBQSR}), which writes a new BAM file with corrected quality scores based on the learned model.

To motivate the need for holistic acceleration of BQSR, Table~\ref{tab:pipeline_breakdown} reports an end-to-end breakdown of the GATK preprocessing pipeline on a whole-genome sequencing (WGS) dataset (D4 from Section~\ref{sec:validation}). As can be seen, the two BQSR phases together account for more than half of the total preprocessing time, making BQSR a high-leverage target for reducing overall workflow latency. Table~\ref{tab:bqsr_breakdown} further decomposes the total runtime of the algorithm into (i) input processing (disk I/O, BGZF decompression, and parsing), (ii) core BQSR computation, and (iii) output processing (formatting, BGZF compression, and writing). The latter is valid only for \texttt{ApplyBQSR}, as \texttt{BaseRecalibrator} does not write out a new BAM file but only a text file containing the recalibration tables, whose I/O cost is negligible, as the tables span only a few KB. The results show that end-to-end performance is constrained not only by compute but also by input/output processing (parsing and (de)compression); \texttt{BaseRecalibrator} is dominated by input handling, whereas \texttt{ApplyBQSR} is dominated by output formatting and compression. 

These measurements motivate the holistic redesign adopted by \textbf{X-BQSR}, where we designed optimized GPU  kernels to handle the complex computation of various stages of BQSR (as described in Section~\ref{subsec:gpu_kernel_design}), and wrapped these kernels in a scalable pipeline that efficiently parallelizes serialization/compression and overlaps computation with I/O to eliminate  bottlenecks in the data path (as described in Section~\ref{subsec:pipeline_optimizations}). 

\subsection{GPU Kernel Design}
\label{subsec:gpu_kernel_design}

\textbf{X-BQSR} contains a series of GPU kernels that implement and adapt the core BQSR algorithms to fully exploit the SIMT parallelism of GPUs. We provide a brief overview of each relevant algorithm in this section together with a detailed description of our SIMT-friendly redesign.

\subsubsection{\texttt{BaseRecalibrator} Sequence clipping}
Sequence clipping identifies adapter regions by scanning the CIGAR (Compact Idiosyncratic Gapped Alignment Report) field. As this scan is a sequential operation over each base of the read, it introduces an intra-read dependency. Hence, we parallelize at the read level by assigning one thread per read so that each thread can maintain its own state while iterating over the read to compute the clipping boundaries. 
Once the region to be clipped has been identified, the sequences must be updated accordingly. This would normally require ``cutting'' the CIGAR, the read itself, and the quality string. All these fields are variable-length strings. To make their representation suitable for the GPU, we store reads using a structure-of-arrays (SoA) layout, where each field across all reads is stored in a large flat vector and the strings of different reads are placed contiguously in memory. For each read, we also maintain a \textit{span vector}  that stores the starting position and length of the read. Instead of costly string manipulation, we implement clipping as a metadata update to this span vector, thus avoiding expensive string copying and manipulation. 

\subsubsection{\texttt{BaseRecalibrator} Skip-slice and SNP Detection}
After sequencing clipping, \texttt{BaseRecalibrator} computes two per-base binary vectors. The first (\texttt{isSNP}) marks sequencing error events by comparing each aligned read to the reference genome. This involves scanning the CIGAR string, advancing read and reference pointers according to edit operators (M, I, D)~\cite{Li2009SAM} and marking each aligned mismatch as an observed error event. The second vector (\texttt{skip mask}) indicates bases that must be excluded from training. This includes positions that (i) overlap known variant sites, (ii) contain ambiguous nucleotides, or (iii) lie outside the post-clipping span. Skip mask is generated by intersecting the read's reference alignment interval with known-site intervals and mapping the resulting locations back to read coordinates.

There are two potential parallelization strategies that can be adopted for computing these two binary vectors on the GPU, namely, a one-thread-per-read mapping or a one-thread-per-base mapping. 

With one-thread-per-read mapping, each thread processes a single read and performs three steps. First, it determines the interval of the reference genome covered by the clipped alignment. Second, it intersects this interval with the list of known variant sites to identify which reference coordinates must be ignored during recalibration. Third, the thread walks the CIGAR operations to translate those reference coordinates into the corresponding read indices. Whenever a reference position associated with a known site maps to a base in the read, the kernel marks that position in the per-read skip mask. SNP events detection is computed similarly, with each thread traversing both read and reference based on the CIGAR and comparing the bases: if they differ, a mismatch is marked in a \texttt{SNP vector}.

The base-parallel design, in contrast, is based on the key idea that a CIGAR string encodes the alignment as a sequence of pairs $<length, operation>$, where the operation can be I (insertion), D (deletion), M (match/mismatch), H (hard clip), and so on. For example, a CIGAR such as $(10M 2D 5M)$ describes ten aligned bases, followed by two deletions and five additional aligned bases. Thus, one possible transformation is to expand this representation so that each operation is repeated per base. In the previous example this would produce a vector such as $MMMMMMMMMMDDMMMMM$. By computing a prefix sum over this expanded vector (reset at the beginning of each read), threads can directly derive the corresponding read and reference coordinates without sequentially walking the CIGAR. This removes the intra-read dependency and produces a more uniform workload across GPU threads.

In \textbf{X-BQSR}, we implemented both these approaches. In our early evaluation, we found read-level parallelism to outperform base-level one as the  transformation requires materializing additional vectors and performing a prefix scan, increasing global memory traffic. Thus, we retained the read-level implementation for this stage.

\subsubsection{\texttt{BaseRecalibrator} Covariate Construction and Table Updates}

Once eligible bases are identified in the previous step, \texttt{BaseRecalibrator} computes the covariates associated with each position. Covariates are discrete features that capture systematic biases in sequencing output. In the \texttt{BaseRecalibrator} formulation, the covariate tuple for base $i$ includes:
\begin{itemize}
    \item Reported quality score $Q_i$: directly extracted from the Phred-encoded quality string; its cardinality is bounded (typically $< 100$).
    \item Cycle covariate $\texttt{cycle}_i$: the base index within the read, computed in a strand-aware manner such that comparable sequencing cycles map to similar values; cardinality is proportional to the maximum read length.
    \item Context covariate $\texttt{context}_i$: a strand-corrected $k$-mer centered at the base (commonly $k=2$ or $3$), encoded in a compact 2-bit representation per base; cardinality is at most $4^k + 1$ (including an invalid sentinel).
\end{itemize}

Concretely, after clipping, masking, and mismatch detection, \texttt{BaseRecalibrator} uses various vectors produced in these stages to compute the covariate tuples, for example, $(RG, Q, cycle, context)$. These tuples represent multidimensional indexes that are used to update tables recording the number of times each covariate appears and the number of mismatches with respect to the reference genome for that base quality.

In the CPU-based GATK \texttt{BaseRecalibrator} implementation, covariate construction is naturally implemented as a single pass where (i) the covariate keys are derived from the eligible bases and (ii) the tables are updated directly. As default GATK \texttt{BaseRecalibrator} is single-threaded, there is no contention when updating the tables. However, when parallelized on the GPU, the tables will be updated simultaneously for thousands of bases. Consequently, updates must be protected with atomics and might experience significant contention. To avoid performance issues, we implemented this stage of \texttt{BaseRecalibrator} using two distinct kernels rather than one.

\paragraph{Covariate emission (base-parallel).}

The first kernel in \textbf{X-BQSR} for this stage just computes covariate indices.
Because this step has no intra-read dependencies, we parallelize it at the base level rather than at the read level.
To do so, we flatten all reads into a single list of bases. Rather than assigning one thread per read, we partition this flattened list into nearly equal contiguous segments that are processed by GPU threads (or groups of threads). Each thread therefore receives approximately the same number of bases to process, independent of the original read boundaries. To determine which read a given base belongs to, we precompute a prefix-sum array over the read lengths. This array stores, for each read, the starting offset of that read in the flattened base list. Given a global base index, a thread performs a search over this prefix array to identify the corresponding read and the offset within that read. Once this mapping is known, the thread can access the appropriate entries in the previously computed vectors (skip mask, mismatch flags, context keys, and quality scores) and compute the covariate indices for that base.
This strategy effectively converts the irregular workload caused by variable-length reads into a balanced distribution of work across GPU threads. As a result, threads within a warp tend to execute a similar amount of work, improving overall utilization of the GPU.

\paragraph{Scalable table aggregation (contention-aware).}

Once the covariate indices and the corresponding per-base SNP information have been computed, the second kernel in \textbf{X-BQSR} updates the covariate tables. Each update consists of indexing the tables with the covariate index and incrementing two values stored at that entry: (i) a counter representing how many times that covariate appears and (ii) a counter tracking the number of mismatches observed for that base.
Since the tables are relatively small while the number of updates is extremely large (billions of bases), concurrent updates on the GPU would result in a very large number of atomic operations that must be serialized due to contention, drastically reducing performance. To mitigate this issue, we adopt a two-level histogram update strategy. Because all tables fit in the GPU shared memory, each thread block first updates a local copy of the tables. Atomic operations within a block are significantly faster than atomics in global memory. Afterward, a single thread per block merges the local histogram into the global table, greatly reducing contention in global memory.

\subsubsection{ApplyBQSR}

Having discussed the design of \textbf{X-BQSR}'s \texttt{BaseRecalibrator} phase, we now focus on the \texttt{ApplyBQSR} phase which rewrites the quality score of each base by evaluating a hierarchical Bayesian model built using the covariate tables computed in the previous phase. Similar to the covariate-emission stage, we use base-level parallelism for redesigning \texttt{ApplyBQSR}.
In particular, for each base, a thread reconstructs the covariate tuple \((\text{readGroup}, \text{qual}, \text{cycle}, \text{context})\), similar to what is done in the \texttt{BaseRecalibrator} algorithm. This tuple is used to retrieve the empirical statistics needed to evaluate the Bayesian model and obtain the recalibrated quality score. 
The resulting quality score is then quantized into the final integer Phred value.

A key observation in this kernel is that many bases may share the same combination of covariates, for example when they have the same quality score, cycle, and context. As a result, the recalibrated quality score may be recomputed repeatedly for identical covariate tuples. To avoid this redundancy, we introduce a device-wide lookup table that stores the result of the computation for each unique covariate combination. The table is implemented as a flattened multidimensional array over the covariate domains (read group, quality score, context, and cycle).
The table size is bounded: up to 94 quality values, roughly 1,000 context values, and up to 1,000 cycle values (500 positions across two strands). Since each entry stores a 1-byte quantized score, the table occupies about 1 MB.
Because BQSR models covariates independently for each read group, the table must be replicated per read group. In practice, however, samples usually contain only 1–10 read groups, so the overall memory footprint remains small.
During execution, each thread computes the covariate indexes and checks whether the recalibrated value is already present in the table. If so, it reuses the cached value directly. Otherwise, it computes the value and atomically updates the table.

\subsection{Pipeline Optimizations}
\label{subsec:pipeline_optimizations}
Having presented the design of GPU kernels in \textbf{X-BQSR}, we now describe various pipeline optimizations that ensure that the acceleration achieved by kernel offload to GPUs is not overshadowed by overheads in the input--output I/O path.

\subsubsection{Optimizing BAM input}
\label{subsubsec:opt_bam_input}

As shown in Table~\ref{tab:bqsr_breakdown}, BAM parsing in input processing was a major overhead for \texttt{BaseRecalibrator}. The primary obstacle to scaling BAM reading is record boundary discovery; after BGZF decompression, the BAM byte stream is a concatenation of variable-length records, and identifying record starts requires a sequential scan. We address this issue by leveraging the \texttt{.bai} index that typically accompanies coordinate-sorted BAM files in best-practice workflows. The BAI linear index provides BGZF virtual offsets that correspond to valid BAM record starts. We extract and deduplicate these virtual offsets and use them as candidate slice boundaries. Because a virtual offset encodes both a compressed block position and an uncompressed offset within the block, it represents a safe restart point for BGZF inflation, enabling deterministic search and decoding.

To balance slices, we build a lightweight BGZF block index recording block offsets and \texttt{ISIZE} values, which lets us estimate decoded slice size and choose the furthest boundary within a target budget. A final tail slice covers all remaining records.

Finally, once the slices are defined, ingestion proceeds via an asynchronous, double-buffered pipeline. Batches correspond to work units transferred to the GPU, which are fixed-size slices that fit within the device's memory. While a batch is parsed and converted to our SoA-based in-memory representation, a separate thread simultaneously reads and decompresses the next batch. This overlap hides the disk I/O and BGZF decompression latency behind the CPU-side parsing and packing work, allowing for sustained throughput to be fed to GPUs.

\subsubsection{Optimizing BAM output}
\label{subsubsec:opt_bam_output}

While the \texttt{BaseRecalibrator} algorithm only generates and writes to disk a small file containing the recalibration tables computed, \texttt{ApplyBQSR} in addition to input and compute stages, must also write a BAM file containing the updated quality score values for each read. The output of a large BAM file can be significant, and it can dominate the end-to-end time of the entire pipeline. This is because it involves several steps: (1) serialization of the BAM record as a BAM record in a flat byte stream; (2) compression of the byte stream in blocks of at most 64 KB; (3) writing the compressed byte blocks (BGZF blocks) to a file.

In order to optimize the BAM output, we exploit multithread CPU parallelism. Each CPU thread  serializes a batch coming from the GPU after quality score recalibration and enqueues the serialized byte buffer (containing the BAM records) for a writer thread to process. It is important to note that the batch can come out-of-order due to different GPUs' processing speeds; therefore, the writer is in charge of processing those batches in order. Once the correct serialized batch arrives, the writer splits it into chunks of 64 KB and assigns each chunk to different threads to be compressed. After compression, each chunk is enqueued to be written on disk.

\subsubsection{Pipelining computation and I/O}
\label{subsubsec:pipeline_io_compute}

Since the pipeline end-to-end throughput is heavily impacted by input and output paths, in order to minimize the latency of the different stages, we built a pipeline where the full input path, the GPU processing stages, and the output path overlap: readers prepare batches asynchronously, GPU worker threads submit them for execution, and completed batches are returned to the host for output processing.
While one batch is executing on the GPU, the next batch is prepared concurrently to keep the device busy.
When the algorithm requires writing a BAM file (as in the case of \texttt{ApplyBQSR}), the batch goes through the output stage: the batches are serialized, compressed in parallel, and written on disk.
\section{Evaluation}
\label{sec:validation}

\definecolor{syclgpu}{HTML}{2E7D32}      
\definecolor{syclcpu}{HTML}{1976D2}      
\definecolor{parabricks}{HTML}{76B900}   
\definecolor{gatkspark}{HTML}{FF6F00}    

\definecolor{nvgreen}{RGB}{46,139,87}
\definecolor{cudablue}{RGB}{0,75,135}
\definecolor{amdred}{RGB}{178,34,34}
\definecolor{intelblue}{RGB}{70,130,180}

In this section, we present the results from our evaluation of \textbf{X-BQSR}. 

\noindent
\textbf{Software Setup.} We have implemented three different variants of \textbf{X-BQSR} using SYCL, CUDA, and HIP in order to assess the performance of vendor-specific implementations. We compare \textbf{X-BQSR} against single-threaded GATK and experimental multi-threaded GATK-Spark v4.6.2.0, NVIDIA Parabricks v4.6.0-1, and an FPGA-accelerated \texttt{BaseRecalibrator} from prior work. All \texttt{X-BQSR} kernels are compiled using the Intel oneAPI DPC++ Compiler v2025.1.0. Native comparison implementations utilize CUDA Toolkit (nvcc) v12.6 for NVIDIA hardware and HIP v6.3.4 (ROCm v6.3.3) for AMD hardware.

\noindent
\textbf{Hardware Setup.} Benchmarks were performed on a dual-socket Intel Xeon Platinum 8454H server (128 logical cores) with 1 TB of RAM. For accelerated workloads, we utilize an NVIDIA L40S (48 GB), an AMD Instinct MI210 (64 GB), and an Intel Flex 170 (16 GB), all via PCIe 4.0. FPGA comparisons utilize a Xilinx VCU118 platform clocked at 100~MHz, connected via PCIe 3.0.

\noindent
\textbf{Datasets.} We use six dataset samples from GIAB individuals (Table~\ref{tab:dataset}): the NA12878 (HG001) and three members of the Ashkenazi trio (HG002, HG003, and HG004). We use the GRCh38 reference for all experiments. After alignment, we process the reads following the GATK Best Practices workflow \cite{GATK42019}. The output of this duplicate-marking step serves as the input for our evaluation. Both \textbf{X-BQSR} and NVIDIA Parabricks produce results identical to the gold-standard single-core GATK.

In all experiments, input and output data reside on local NVMe storage to ensure consistent I/O performance. All experiments were conducted 10 times to ensure statistical rigor; the reported values represent the average of these runs, and because the results exhibited a negligible deviation of less than 5\%, error bars are omitted from the figures for clarity.

\begin{table}[t]
\centering
\caption{Datasets used for validation}
\label{tab:dataset}
\begin{tabular}{lrrrr}
\toprule
\textbf{Dataset} & \textbf{Name} & \textbf{\#Reads} & \textbf{Read Len} \\
\midrule
D0 & HG001-HiSeq2500 & 168.544.418 & 100 \\
D1 & HG001-HiSeq4000 & 88.774.172 & 150 \\
D2 & HG002-HiSeqX & 777.447.412 & 150 \\
D3 & HG003-HiSeqX & 1.001.580.652 & 150 \\
D4 & HG004-NovaSeq & 1.166.740.818 & 150 \\
D5 & HG002-HiSeq2500 & 898.865.538 & 250 \\
\bottomrule
\end{tabular}
\vspace{-5mm}
\end{table}

\subsection{Comparison with State-of-the-Art}

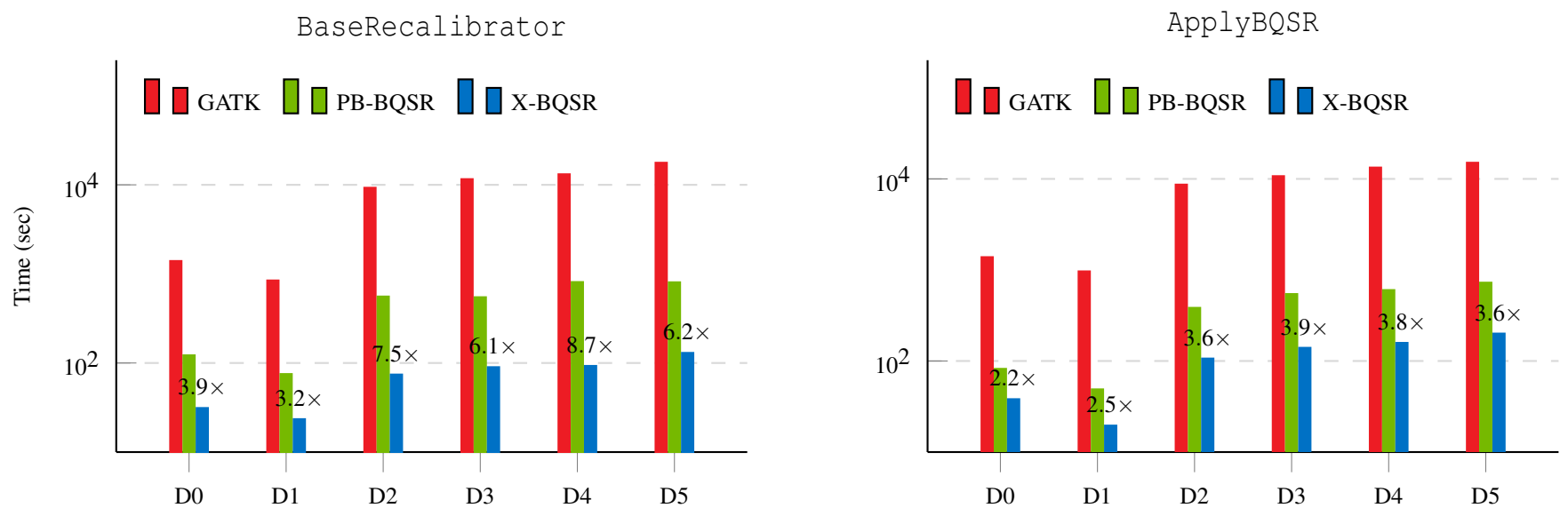
\begin{figure*}[t]
\centering
\pgfplotsset{
    compat=1.18, 
    bqsr_style/.style={
        ybar=0pt, 
        width=1.05\linewidth, 
        height=4.5cm,        
        bar width=2.8pt,     
        ymajorgrids=true,
        grid style={dashed, gray!30},
        tick label style={font=\tiny},
        title style={font=\scriptsize, yshift=-1ex},
        label style={font=\tiny},
        axis lines*=left,
        symbolic x coords={D0, D1, D2, D3, D4, D5},
        xtick=data,
        enlarge x limits=0.15, 
        ymin=10,
        enlarge y limits={upper, value=0.35},
    }
}

\begin{subfigure}[t]{0.49\linewidth}
\centering
\begin{tikzpicture}
\begin{semilogyaxis}[
    bqsr_style,
    log basis y=10,
    ylabel={Time (sec)},
    title={\texttt{BaseRecalibrator}},
    legend style={
        at={(0.02,0.98)}, 
        anchor=north west,
        legend columns=3,
        draw=none,
        fill=none,
        font=\tiny, 
        /tikz/every even column/.append style={column sep=0.3em}
    }
]
\addplot[fill=gatkred, draw=none] coordinates {(D0, 1430) (D1, 865) (D2, 9500) (D3, 11869) (D4, 13461) (D5, 18134)};
\addplot[fill=parabricksgreen, draw=none] coordinates {(D0, 125) (D1, 77) (D2, 571) (D3, 561) (D4, 830) (D5, 825)};
\addplot[fill=syclgalblue, draw=none, nodes near coords, point meta=explicit symbolic, every node near coord/.append style={font=\tiny, anchor=south, yshift=-1pt}] 
    coordinates {(D0, 32) [$3.9\times$] (D1, 24) [$3.2\times$] (D2, 76) [$7.5\times$] (D3, 92) [$6.1\times$] (D4, 95) [$8.7\times$] (D5, 133) [$6.2\times$]};
\legend{GATK, PB-BQSR, X-BQSR}
\end{semilogyaxis}
\end{tikzpicture}
\end{subfigure}
\hfill 
\begin{subfigure}[t]{0.49\linewidth}
\centering
\begin{tikzpicture}
\begin{semilogyaxis}[
    bqsr_style,
    log basis y=10,
    title={\texttt{ApplyBQSR}},
    ylabel={}, 
    legend style={ 
        at={(0.02,0.98)}, 
        anchor=north west,
        legend columns=3,
        draw=none,
        fill=none,
        font=\tiny, 
        /tikz/every even column/.append style={column sep=0.3em}
    }
]
\addplot[fill=gatkred, draw=none] coordinates {(D0, 1416) (D1, 988) (D2, 8869) (D3, 10983) (D4, 13651) (D5, 15424)};
\addplot[fill=parabricksgreen, draw=none] coordinates {(D0, 84) (D1, 50) (D2, 393) (D3, 557) (D4, 616) (D5, 743)};
\addplot[fill=syclgalblue, draw=none, nodes near coords, point meta=explicit symbolic, every node near coord/.append style={font=\tiny, anchor=south, yshift=-1pt}] 
    coordinates {(D0, 39) [$2.2\times$] (D1, 20) [$2.5\times$] (D2, 109) [$3.6\times$] (D3, 143) [$3.9\times$] (D4, 162) [$3.8\times$] (D5, 205) [$3.6\times$]};
\legend{GATK, PB-BQSR, X-BQSR}
\end{semilogyaxis}
\end{tikzpicture}
\end{subfigure}

\caption{Execution time comparison of \textbf{X-BQSR} against GATK and NVIDIA Parabricks (D0--D5). Speedup values ($\times$) are relative to NVIDIA Parabricks.}
\label{fig:bqsr_gatk_vs_accel}
\vspace{-5mm}
\end{figure*}

We start the evaluation by benchmarking \texttt{X-BQSR} against the standard single-threaded GATK implementation.
The execution results for the \texttt{BaseRecalibrator} and \texttt{ApplyBQSR} stages (Figure~\ref{fig:bqsr_gatk_vs_accel}) clearly show the performance gap between traditional CPU methods and accelerated implementations. The single-threaded GATK baseline serves as the upper bound for execution time, requiring over 5 hours (18,134 seconds) for the D5 dataset. Across the evaluated production-scale datasets (D2--D5), \textbf{X-BQSR} \texttt{BaseRecalibrator} achieves speedups over the GATK gold standard of 125.0$\times$ (D2), 129.0$\times$ (D3), 141.7$\times$ (D4), and 136.3$\times$ (D5). For \texttt{ApplyBQSR}, \textbf{X-BQSR} achieves improvement ranging from 75.2$\times$ to 84.3$\times$ over single-threaded CPU execution.

In the GPU-accelerated category, we compare \textbf{X-BQSR} with the proprietary NVIDIA Parabricks-BQSR tool.  In \texttt{BaseRecalibrator}, \textbf{X-BQSR} consistently outperforms Parabricks with speedups ranging from $6.1\times$ to $8.7\times$ for the larger  datasets (D2–D5). For instance, on the D4 dataset, \textbf{X-BQSR} completes the stage in 95 seconds compared to 830 seconds for Parabricks. For  \texttt{ApplyBQSR}, \textbf{X-BQSR} exhibits a consistent $3.6\times$ to $3.9\times$ speedup over Parabricks across the D2–D5 datasets. Specifically, \textbf{X-BQSR} processes the 1.16 billion reads of D4 in 162 seconds (vs. 616 seconds for Parabricks) and the 898 million reads of D5 in 205 seconds (vs. 743 seconds for Parabricks). As Parabricks is provided as a precompiled binary within a Docker container, its optimization strategies and kernel configurations are opaque. Thus, it is not possible for us to identify the specific causes of the performance gap with Parabricks. In any case, these results demonstrate that our open, standards-based solution outperforms state-of-the-art proprietary ones.

\subsection{Multi-GPU and CPU Scalability}

\begin{figure*}[t]
\centering
\begin{subfigure}{0.33\textwidth}
\centering
\begin{tikzpicture}
\begin{axis}[
    ybar, width=1.1\linewidth, height=4.5cm,
    bar width=9pt, ylabel={\tiny Time (sec)},
    symbolic x coords={1 GPU, 2 GPUs}, xtick=data,
    tick label style={font=\tiny}, ymajorgrids=true,
    grid style={dashed, gray!30}, 
    ymin=0, axis lines*=left, 
    enlarge x limits=0.5,
    legend style={at={(0.45, 0.95)}, anchor=north, draw=none, fill=none, font=\tiny, legend columns=1}
]
\addplot[fill=parabricksgreen, draw=none] coordinates { (1 GPU, 817) (2 GPUs, 815) };
\addplot[fill=syclgalblue, draw=none] coordinates { (1 GPU, 220) (2 GPUs, 134) };
\legend{PB-BQSR, X-BQSR}
\end{axis}
\end{tikzpicture}
\caption{GPU \texttt{BaseRecalibrator}}
\label{subfig:gpu_recal_scalability}
\end{subfigure}\hspace{-2mm} 
\begin{subfigure}{0.33\textwidth}
\centering
\begin{tikzpicture}
\begin{axis}[
    ybar, width=1.1\linewidth, height=4.5cm,
    bar width=9pt, 
    symbolic x coords={1 GPU, 2 GPUs}, xtick=data,
    tick label style={font=\tiny}, ymajorgrids=true,
    grid style={dashed, gray!30},
    ymin=0, axis lines*=left, 
    enlarge x limits=0.5,
    legend style={at={(0.45, 0.95)}, anchor=north, draw=none, fill=none, font=\tiny, legend columns=1}
]
\addplot[fill=parabricksgreen, draw=none] coordinates { (1 GPU, 693) (2 GPUs, 693) };
\addplot[fill=syclgalblue, draw=none] coordinates { (1 GPU, 234) (2 GPUs, 213) };
\legend{PB-BQSR, X-BQSR}
\end{axis}
\end{tikzpicture}
\caption{GPU \texttt{ApplyBQSR}}
\label{subfig:gpu_apply_scalability}
\end{subfigure}
\begin{subfigure}{0.33\textwidth}
\centering
\begin{tikzpicture}
\begin{axis}[
    width=1.1\linewidth, height=4.5cm,
    xlabel={\tiny Threads}, 
    symbolic x coords={2,4,8,16,32,64,128},
    xtick=data, tick label style={font=\tiny},
    grid=major, grid style={dashed, gray!30},
    ymin=0, axis lines*=left,
    scaled y ticks=base 10:4,
    ytick scale label code/.code={
        \node[anchor=south west, inner sep=0pt, font=\tiny, yshift=-1.5em] at (rel axis cs:0,1) {$\times 10^4$};
    },
    legend style={at={(0.98,0.98)}, anchor=north east, draw=none, fill=white, fill opacity=0.8, font=\tiny}
]
\addplot[color=blue!70!black, mark=*, mark size=1pt] coordinates {
    (2, 8490) (4, 4467) (8, 2314) (16, 1238) (32, 715) (64, 464) (128, 411)
};
\addplot[color=orange!85!black, mark=square*, mark size=1pt] coordinates {
    (2, 17315) (4, 8591) (8, 4512) (16, 2388) (32, 1461) (64, 1156) (128, 1226)
};
\legend{\texttt{BaseRecalibrator}, \texttt{ApplyBQSR}}
\end{axis}
\end{tikzpicture}
\caption{GATK-Spark CPU Scaling}
\label{subfig:cpu_scalability}
\end{subfigure}

\caption{Scaling D5 dataset across GPU (\textbf{X-BQSR}, Parabricks) and CPU (GATK-Spark)}
\label{fig:combined_scalability}
\vspace{-5mm}
\end{figure*}
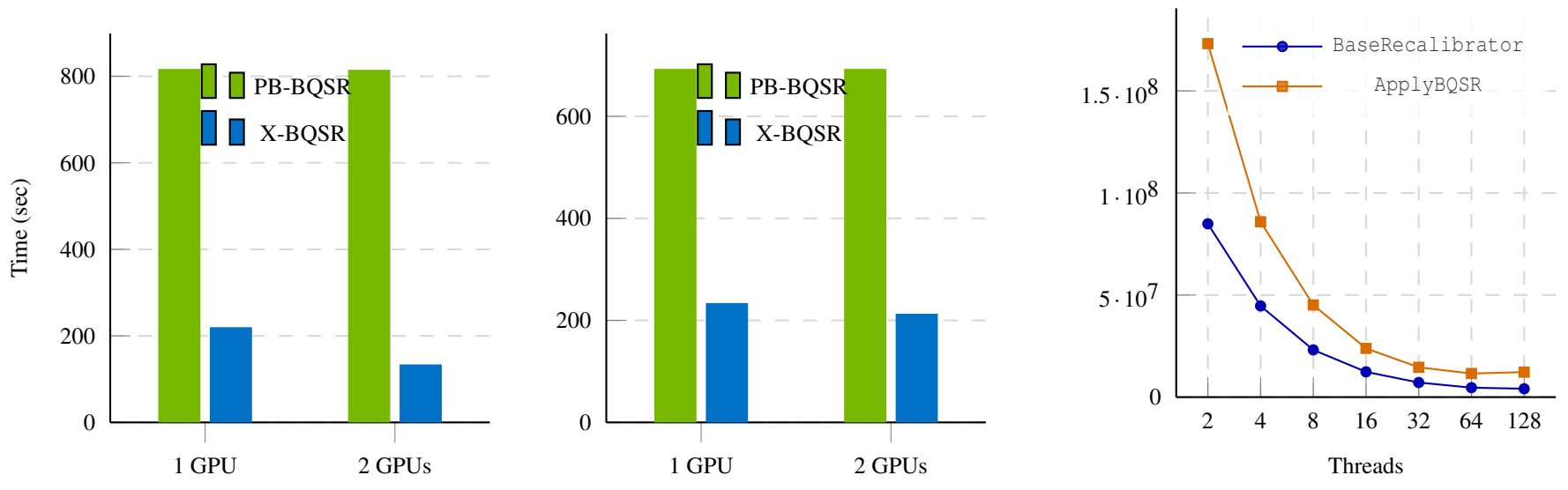

Beyond single-device performance, we evaluate the scalability of our implementation by distributing the BQSR workload across multiple accelerators using dataset D5. As shown in Figures~\ref{subfig:gpu_recal_scalability} and~\ref{subfig:gpu_apply_scalability}, \textbf{X-BQSR} demonstrates superior scaling compared to Parabricks. In \texttt{BaseRecalibrator}, \textbf{X-BQSR} reduces execution time from 220 seconds to 134 seconds when moving from one to two GPUs, whereas Parabricks shows almost no improvement (815s vs. 813s). In the \texttt{ApplyBQSR} stage, achieving linear scaling is more difficult due to the requirement of a single-threaded writer process that must serialize data to disk in a specific order to maintain GATK compatibility. As the number of GPUs increases, the computational portion of the workload is reduced, but this ordering requirement  leads to output I/O becoming a dominating time sink, effectively capping the maximum achievable speedup.

GATK-Spark is an experimental, multi-threaded implementation of BQSR based on Apache Spark. Unlike the original GATK BQSR implementation, the Spark version, to our knowledge, is not widely used. However, we show the scaling results here for completeness. Figure~\ref{subfig:cpu_scalability} illustrates the scalability of GATK-Spark from 2 to 128 threads on the D5 dataset. While multi-threading significantly reduces execution times compared to the single-threaded baseline—bringing \texttt{BaseRecalibrator} down to 411 seconds and \texttt{ApplyBQSR} to 1,226 seconds at 128 threads, it remains significantly slower than a single GPU. Specifically, when utilizing all 128 threads of the dual-socket server, the GATK-Spark \texttt{BaseRecalibrator} stage is 3.1$\times$ slower (411s vs. 133s) and the \texttt{ApplyBQSR} stage is 6.0$\times$ slower (1,226s vs. 205s) than a single NVIDIA L40S running \textbf{X-BQSR}.

\subsection{Kernel Performance and Cross-Architecture Portability}
\label{subsec:cross-arch-portability}

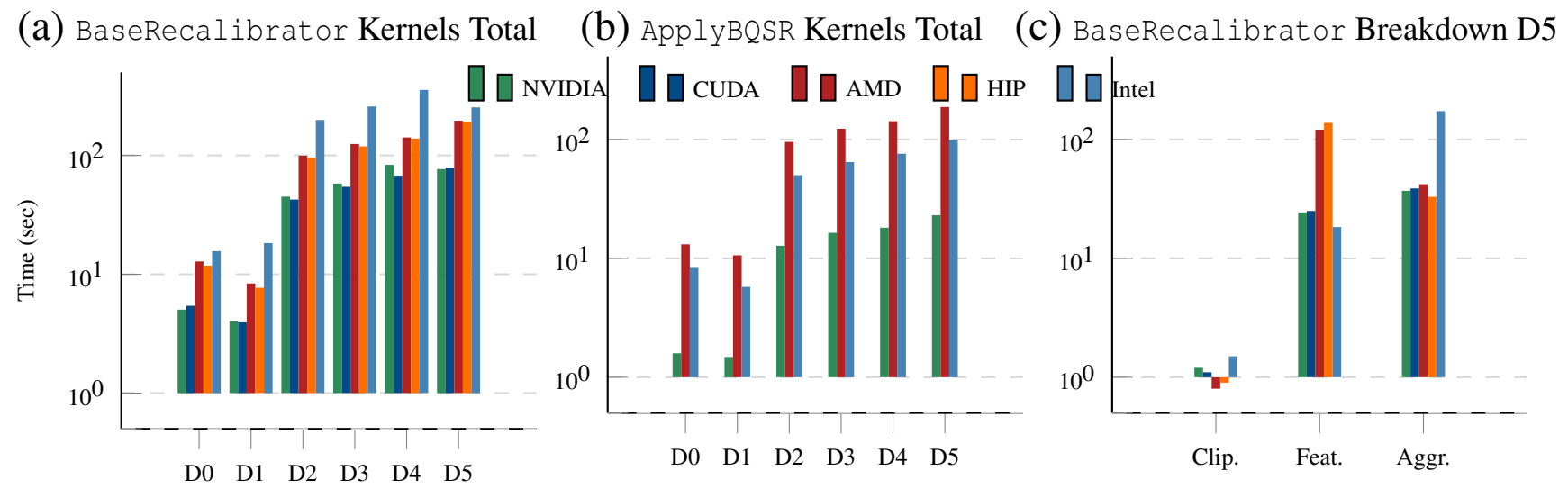
\begin{figure*}[t]
\centering
\pgfplotsset{
    compact_style/.style={
        ybar=0pt, 
        height=4.2cm,
        bar width=1.8pt, 
        ymajorgrids=true,
        grid style={dashed, gray!30},
        tick label style={font=\tiny},
        title style={font=\scriptsize, yshift=-1ex},
        axis lines*=left,
        enlarge x limits=0.3,
        xtick=data,
        ytick={1, 10, 100},
        extra y ticks={0.5},
        extra y tick labels={}, 
    }
}

\begin{subfigure}[t]{0.33\textwidth}
\caption{\scriptsize \texttt{BaseRecalibrator} Kernels Total}
\centering
\begin{tikzpicture}
\begin{axis}[
    compact_style,
    ymode=log,
    ymin=0.5,
    ymax=500,
    log basis y=10,
    width=1.15\linewidth,
    ylabel={\tiny Time (sec)},
    symbolic x coords={D0, D1, D2, D3, D4, D5},
    legend style={at={(1.675,1.05)}, anchor=north, legend columns=5, draw=none, fill=none, font=\tiny, /tikz/every even column/.append style={column sep=0.5em}}
]
\addplot[fill=nvgreen, draw=none] coordinates {(D0, 5.04) (D1, 4.03) (D2, 45.02) (D3, 57.96) (D4, 83.28) (D5, 76.78)};
\addplot[fill=cudablue, draw=none] coordinates {(D0, 5.43) (D1, 3.94) (D2, 42.54) (D3, 54.44) (D4, 67.59) (D5, 78.99)};
\addplot[fill=amdred, draw=none] coordinates {(D0, 12.82) (D1, 8.37) (D2, 99.54) (D3, 124.82) (D4, 141.81) (D5, 195.98)};
\addplot[fill=gatkspark, draw=none] coordinates {(D0, 11.82) (D1, 7.70) (D2, 95.91) (D3, 119.10) (D4, 138.60) (D5, 191.35)};
\addplot[fill=intelblue, draw=none] coordinates {(D0, 15.65) (D1, 18.31) (D2, 198.65) (D3, 258.15) (D4, 356.12) (D5, 253.92)};
\legend{NVIDIA, CUDA, AMD, HIP, Intel}
\end{axis}
\end{tikzpicture}
\end{subfigure}\hspace{-4mm}
\begin{subfigure}[t]{0.33\textwidth}
\caption{\scriptsize \texttt{ApplyBQSR} Kernels Total}
\centering
\begin{tikzpicture}
\begin{axis}[
    compact_style,
    ymode=log,
    ymin=0.5,
    ymax=500,
    log basis y=10,
    width=1.15\linewidth,
    symbolic x coords={D0, D1, D2, D3, D4, D5},
    ylabel={}, 
]
\addplot[fill=nvgreen, draw=none] coordinates {(D0, 1.59) (D1, 1.48) (D2, 12.77) (D3, 16.40) (D4, 18.12) (D5, 23.05)};
\addplot[fill=amdred, draw=none] coordinates {(D0, 13.12) (D1, 10.61) (D2, 95.51) (D3, 123.27) (D4, 142.64) (D5, 187.82)};
\addplot[fill=intelblue, draw=none] coordinates {(D0, 8.32) (D1, 5.75) (D2, 50.10) (D3, 64.54) (D4, 75.92) (D5, 99.27)};
\end{axis}
\end{tikzpicture}
\end{subfigure}\hspace{-4mm}
\begin{subfigure}[t]{0.33\textwidth}
\caption{\scriptsize \texttt{BaseRecalibrator} Breakdown D5}
\centering
\begin{tikzpicture}
\begin{axis}[
    compact_style,
    ymode=log,
    ymin=0.5,
    ymax=500,
    log basis y=10,
    width=1.15\linewidth,
    symbolic x coords={Clip, Feature, Aggr},
    xticklabels={Clip., Feat., Aggr.},
    enlarge x limits=0.5,
    ylabel={},
]
\addplot[fill=nvgreen, draw=none] coordinates {(Clip, 1.2) (Feature, 24.37) (Aggr, 37.02)};
\addplot[fill=cudablue, draw=none] coordinates {(Clip, 1.1) (Feature, 25.06) (Aggr, 38.76)};
\addplot[fill=amdred, draw=none] coordinates {(Clip, 0.8) (Feature, 121.07) (Aggr, 42.06)};
\addplot[fill=gatkspark, draw=none] coordinates {(Clip, 0.9) (Feature, 137.99) (Aggr, 32.95)};
\addplot[fill=intelblue, draw=none] coordinates {(Clip, 1.5) (Feature, 18.34) (Aggr, 173.33)};
\end{axis}
\end{tikzpicture}
\end{subfigure}

\caption{Cross-architecture kernel performance comparison}
\label{fig:performance_summary}
\vspace{-5mm}
\end{figure*}

In this section, we present a study of performance portability of \textbf{X-BQSR} kernels on multi-vendor GPUs. By focusing on kernel-only execution time (Figure~\ref{fig:performance_summary}a and \ref{fig:performance_summary}b), we isolate the algorithmic performance from host-side I/O and driver overhead. As can be seen, the SYCL-based \textbf{X-BQSR} kernels are highly competitive with native toolchains. On the D5 dataset, our \texttt{BaseRecalibrator} kernels on the NVIDIA backend outperform optimized native CUDA kernels by approximately 2.8\% (76.78~s vs. 78.99~s). For AMD hardware, while our initial SYCL port demonstrated portability, the native HIP implementation provides a measurable speedup, reducing the total \texttt{BaseRecalibrator} kernel time for D5 from 195.98~s to 191.35~s. This native implementation leverages AMD-specific compiler optimizations and avoids the intermediate abstraction layers of the SYCL runtime. In the \texttt{ApplyBQSR} phase, the performance hierarchy aligns with raw hardware specifications, with the NVIDIA L40S leading, followed by the Intel Flex 170 and the AMD MI210.

Figure~\ref{fig:performance_summary}c presents the breakdown of execution time across various stages of \textbf{X-BQSR} to identify specific architectural bottlenecks. In \texttt{BaseRecalibrator}, the \textbf{Feature Derivation} stage, which encompasses skip-masking, SNP detection, and context covariate emission, is the primary execution driver on AMD hardware. While we ensure high GPU occupancy, the complex instruction mix and address generation required for context emission become compute-limited on the CDNA2 SIMD units.

Conversely, the \textbf{Table Aggregation} stage highlights the efficacy of our two-level aggregation strategy. By utilizing shared-memory histograms to minimize global atomic contention, architectures with high L2-cache atomic resolution, like the AMD MI210, remain highly competitive. Intel, however, shows significant overhead in this stage, as the Xe-HPG architecture resolves atomics with higher latency, confirming that while our design successfully mitigates hardware contention, performance remains bounded by the underlying atomic throughput of the GPU.

To further understand the architectural drivers behind the performance trends observed across multi-vendor GPUs, we use two SYCL microbenchmarks to isolate the hardware limits of BQSR: (i) the contended atomics benchmark (mimicking \textbf{Table Aggregation}) and (ii) the STREAM triad benchmark (mimicking the \texttt{ApplyBQSR} phase streaming pattern). The contended-atomics benchmark performs $R$ \texttt{fetch\_add} operations into a global table of 94 bins using \texttt{sycl::atomic\_ref}. The 94-bin size is chosen to match the \textbf{Cycle} covariate, which is the primary source of atomic contention in \texttt{BaseRecalibrator}. The STREAM triad benchmark measures steady-state device bandwidth, mirroring the memory-access pattern of the \texttt{ApplyBQSR} phase.

\begin{table}[t]
\centering
\small
\caption{Atomic update throughput (Gop/s) and sustained device bandwidth (GB/s)}
\label{tab:microbench}
\setlength{\tabcolsep}{6pt} 
\renewcommand{\arraystretch}{1.15}
\begin{tabular}{l c c} 
\toprule
Device & Contended Atomics (Gop/s) & STREAM (GB/s) \\
\midrule
NVIDIA L40S    & 3.09  & 689  \\
AMD MI210      & 11.37 & 1302 \\
Intel Flex 170 & 0.90  & 344  \\
\bottomrule
\end{tabular}
\vspace{-5mm}
\end{table}

Table~\ref{tab:microbench} presents the results from these benchmarks. The AMD MI210 reaches 11.37~Gop/s on atomics, $4\times$ higher than the NVIDIA L40S. Hardware counters showed that $>$99\% of these operations are resolved within the MI210's L2 cache.
This explains why AMD remains highly competitive in the \textbf{Table Aggregation} stage despite having lower raw compute clock speeds than the L40S.
For memory bandwidth, the MI210 sustains 1302~GB/s, while the L40S sustains 689~GB/s. Despite this $1.9\times$ bandwidth advantage, the \texttt{ApplyBQSR} phase is substantially slower on AMD. Profiling shows high VALU utilization on the MI210, suggesting that the SYCL instruction mix for address generation and hierarchical Bayesian calculation becomes compute-limited on the CDNA2 SIMD units, preventing the kernels from saturating the available HBM2e bandwidth. 
The Intel Flex 170 sustains 344~GB/s but is limited by its lower atomic throughput (0.90~Gop/s), which is consistent with its longer \textbf{Table Aggregation} times.
A quantitative analysis confirms the predictive value of these metrics; while atomic throughput successfully predicts aggregation rankings, the \texttt{ApplyBQSR} performance on AMD highlights the critical importance of compiler-level instruction scheduling in cross-architecture SYCL deployments.
\vspace{-5mm}
\subsection{FPGA Comparison}
\label{sec:eval-fpga}

\begin{table*}[t]
    \centering
    \small 
    \caption{BQSR execution time and speedup relative to the GATK single-core baseline}
    \label{tab:bqsr_comparison}
    
    \setlength{\tabcolsep}{3pt} 
    
    \begin{tabular*}{\textwidth}{@{\extracolsep{\fill}} l r r r r r r @{}}
        \toprule
        \textbf{Data} & 
        \textbf{\makecell{GATK\\(s)}} & 
        \textbf{\makecell{Extract\\(s)}} & 
        \textbf{\makecell{Construct\\(s)}} & 
        \textbf{\makecell{FPGA-BQSR\\(s / Speedup)}} & 
        \textbf{\makecell{X-BQSR FPGA\\(s / Speedup)}} & 
        \textbf{\makecell{X-BQSR GPU\\(s / Speedup)}} \\
        \midrule
        1M  & 16.20  & 50.77  & 22.68   & 0.70 / 23.2$\times$  & 12.80 / 1.27$\times$ & 0.05 / 341.3$\times$ \\ 
        2M  & 30.00  & 82.57  & 43.54   & 1.33 / 23.5$\times$  & 13.12 / 2.29$\times$ & 0.10 / 300.0$\times$ \\ 
        4M  & 60.00  & 163.09 & 85.95   & 2.60 / 23.1$\times$  & 12.91 / 4.65$\times$ & 0.21 / 292.5$\times$ \\ 
        8M  & 126.00 & 336.77 & 494.99  & 5.11 / 24.6$\times$  & 13.77 / 9.15$\times$ & 0.42 / 303.6$\times$ \\ 
        16M & 246.00 & 721.07 & 1,133.82 & 10.15 / 24.2$\times$ & 15.12 / 16.3$\times$ & 0.83 / 295.3$\times$ \\ 
        \bottomrule
    \end{tabular*}
    \vspace{-5mm}
\end{table*}

To further contextualize our performance, we compare \textbf{X-BQSR} against a state-of-the-art FPGA redesign of BQSR by Lo et al., which provides a solution for accelerating only the \texttt{BaseRecalibrator} stage of BQSR. They do not focus on the second phase of BQSR (\texttt{ApplyBQSR}). We present a comparison of our GPU-accelerated \texttt{BaseRecalibrator} in \textbf{X-BQSR} with Lo et al's FPGA solution for completeness. With the FPGA accelerator architecture, the host partitions the dataset into read chunks of size 100,000 and sends them to the FPGA sequentially for writing into SDRAM. Due to limited FPGA resources, we cannot process all 100,000 reads in a single iteration. Only 16 processing elements are implemented on the FPGA, so all 100,000 reads are processed across 6,250 iterations. Upon completing a chunk of reads, the BQSR accelerator asserts an interrupt to notify the host to send the next chunk. To maximize throughput, the host transfers the next chunk of reads to DRAM concurrently while the FPGA processes the previous chunk.

Unlike our GPU implementation, the FPGA architecture requires data to be preformatted into a specific binary structure. Consequently, an end-to-end FPGA workflow necessitates two additional pre-processing phases: (i) the extraction of bases, quality scores, and error metadata from the input BAM file, and (ii) the construction of the FPGA-specific data structures. Table~\ref{tab:bqsr_comparison} summarizes the execution times for the three phases of the FPGA workflow alongside the GATK baseline. While the standalone FPGA-BQSR compute kernel achieves a speedup of up to $24.6\times$ over GATK, this is lower than the $40\times$ speedup reported in the original study. Our profiling indicates that the discrepancy stems from memory latency: the Load stage (fetching data from FPGA DRAM) requires 865 clock cycles in our environment, compared to the 534 cycles reported by Lo et al. using a 4-channel DRAM configuration. At 100~MHz, this results in a BQSR execution time of approximately 0.054~s per 100,000 reads ($865 \times 6,250 \times 10 \times 10^{-9} \approx 0.054$~s).

The primary limitation of the FPGA approach is revealed when considering end-to-end integration. The required pre-processing phases (Extract and Construct) take $4\times$ to $7\times$ longer than the entire single-threaded GATK execution, effectively negating the gains of the hardware accelerator. In contrast, our \textbf{X-BQSR} implementation provides a holistic redesign that accelerates both I/O and computation. By eliminating hidden pre-computation overheads, \textbf{X-BQSR} serves as a seamless drop-in replacement for GATK, providing true end-to-end acceleration for germline variant calling pipelines.
\vspace{-6mm}
\section{Conclusion}

In this work, we focused on BQSR, one of the most time-consuming steps of the widely used GATK germline variant calling pipeline. We motivated the need for hardware acceleration of BQSR and showed how focusing only on GPU acceleration of the computational part, as done by prior work, will only yield marginal improvement in overall time due to I/O bottlenecks. We presented \textbf{X-BQSR}, our holistic redesign with several I/O and compute optimizations, and showed how \textbf{X-BQSR} can provide up to 141.7$\times$ improvement over GATK and up to 8.7$\times$ improvement over closed-source solutions while being vendor-agnostic, open source\footnote{\url{https://gitlab.eurecom.fr/dislab/bqsr}}, and performance portable. In future work, we first plan to integrate \textbf{X-BQSR} with other preprocessing stages (alignment, sorting and mark duplicates) and then with downstream variant callers to develop a fully open, standards-based, GPU-accelerated solution for scaling genomic data analysis.

\bibliographystyle{splncs04}
{\footnotesize
\bibliography{Sections/ref}}

\end{document}